# Near-elliptic core triangular-lattice and square-lattice PCFs: a comparison of birefringence, cut-off and GVD characteristics towards fiber device application

Partha Sona Maji* and Partha Roy Chaudhuri

Department of Physics & Meteorology, Indian Institute of Technology Kharagpur-721 302, INDIA
*Tel: +91-3222-283842 Fax: +91-3222-255303,*
*Corresponding author: parthamaji@phy.iitkgp.ernet.in*

**Abstract:** In this work, we report detailed numerical analysis of the near-elliptic core index-guiding triangular-lattice and square-lattice photonic crystal fiber (PCFs); where we numerically characterize the birefringence, single mode, cut-off behavior and group velocity dispersion and effective area properties. By varying geometry and examining the modal field profile we find that for the same relative values of $d/\Lambda$, triangular-lattice PCFs show higher birefringence whereas the square-lattice PCFs show a wider range of single-mode operation. Square-lattice PCF was found to be endlessly single-mode for higher air-filling fraction ($d/\Lambda$). Dispersion comparison between the two structures reveal that we need smaller lengths of triangular-lattice PCF for dispersion compensation whereas PCFs with square-lattice with nearer relative dispersion slope (RDS) can better compensate the broadband dispersion. Square-lattice PCFs show zero dispersion wavelength (ZDW) red-shifted, making it preferable for mid-IR supercontinuum generation (SCG) with highly non-linear chalcogenide material. Square-lattice PCFs show higher dispersion slope that leads to compression of the broadband, thus accumulating more power in the pulse. On the other hand, triangular-lattice PCF with flat dispersion profile can generate broader SCG. Square-lattice PCF with low Group Velocity Dispersion (GVD) at the anomalous dispersion corresponds to higher dispersion length ($L_D$) and higher degree of solitonic interaction. The effective area of square-lattice PCF is always greater than its triangular-lattice counterpart making it better suited for high power applications. We have also performed a comparison of the dispersion properties of between the symmetric-core and asymmetric-core triangular-lattice PCF. While we need smaller length of symmetric-core PCF for dispersion compensation, broadband dispersion compensation can be performed with asymmetric-core PCF. Mid-Infrared (IR) SCG can be better performed with asymmetric core PCF with compressed and high power pulse, while wider range of SCG can be performed with symmetric core PCF. Thus, this study will be extremely useful for designing/realizing fiber towards a custom application around these characteristics.

## 1. INTRODUCTION

Photonic Crystal Fiber (PCFs)[1-2], which offers a great flexibility to vary the design parameters (*i.e.*, air-hole diameter "*d*" and hole-to-hole separation Λ) at the fabrication stage, provides the freedom to design the PCF for a number of unique and useful applications like dispersion tailoring, wideband single-mode operation, controlling losses, achieving higher effective area for handling high power applications like supercontinuum generation etc. which are not achievable in standard step index silica fibers. Importantly, with their remarkable and extraordinary features in light-guiding, PCFs are found to be absolutely unique in many applications in the field of photonics. Well-matured technology is now developed in telecommunication and signal processing applications, metrology, instrumentation, medical surgery and imaging spectroscopy [3].

Some of the unusual and exciting characteristics that have revolutionized the light-guiding features of PCFs are high-birefringence [4-15], endlessly single-mode operation [16], high effective nonlinearity [17] and tailoring of Group Velocity Dispersion (GVD) [18]. Essentially these characteristics dictate most of today's fiber-based device applications, in particular the nonlinear fiber devices. For example, a proper study of highly birefringent fiber is extremely necessary for its many applications such as sensing [19-21], long period grating devices [22], fiber ring laser [23], phase matching for four-wave mixing [24], parametric amplifier [25-28] etc. In PCF, a very high index contrast is available between silica and air-holes. This property, when combined with a large anisotropic geometry, will produce a huge amount of birefringence. For this purpose of anisotropy geometry, several approaches have been followed such as altering the associated structural parameters namely, asymetric core [10], elliptical air-holes [9], different sized air-holes [5, 13] and air-holes filling with selective liquids along one of their diagonals [11, 15]. Out of the above methods, asymmetric core PCF is the most studied one. Asymmetric core PCF can be created with two adjacent air-holes missing in the core region. The concept has been extensively studied and demonstrated for triangular-lattice PCF [8, 10]. Similar perception can be applicable for square-lattice PCF with "Λ" as the hole-to-hole spacing both in horizontal and vertical directions in the PCF and "*d*" as the diameter of the air holes. To realize asymmetric core square-lattice PCF, two adjacent air-holes can be removed from the core region to induce the ellipticity. Symmetric core PCFs are termed as endlessly single-mode up-to a certain $d/\Lambda$. The values are 0.406 for a triangular one and 0.446 for a square-lattice one. The above values of cut-off between single-mode and multimode regions cannot be the same for elliptical core PCF. The exact value of the cut-off between single mode and multimode is an important parameter for applications like high power laser delivery in single-mode regime [29-32], to maximize the overlap amongst signal power, pump power and doped region at the core of an amplifier fiber [33-34]. The dispersion properties of this fiber; especially the tailoring of zero dispersion wavelength (ZDW) is extremely useful for numerous applications like broad-band supercontinuum generation (SCG) [35], ASE (Amplified Spontaneous Emission) suppressed amplifier [34], reduction of soliton fission in SCG [36] and different sources for Infrared (IR) applications [37]. The above features of PCF are primarily owed to the photonic lattice structure of the fiber. There are two competing technologies available in realizing these microstructured fibers, the triangular-lattice [1-2, 18] and square-lattice transverse geometry [38-39]. Both the structural approaches have certain advantages and disadvantages. In view of these facts, in this paper, we describe our investigations on the comparative assessment of these key modal properties, namely, birefringence, cut-off and dispersion characteristics of the asymmetric core PCFs with triangular-lattice and square-lattice structure with $C_2v$ symmetry. $C_n$ symmetry implies that the structure remains unchanged after a rotation of $2\pi/n$. The symbol $C_n$ stands for both a particular symmetry operation and the collection of all symmetry operations based on it. The term $C_nv$ means that the structure also includes $n$ number of planes of reflection symmetry along with the $n$ fold rotational symmetry. For our case $n=2$. A detail of the symmetry can be obtained from McIssac [40].

Figures 1(a) and (b) respectively show the triangular-lattice and square-lattice cladding structures with two adjacent central holes missing in a silica matrix background. In the analysis of both the structures, we use the conventional notations of air-hole diameter as "$d$" and hole-to-hole distance as "$\Lambda$". However, we use $\Lambda$ as the hole-to-hole spacing both in horizontal and vertical direction in the PCF with square-lattice air holes and $d$ as the diameters of the air holes. The modal field, the dispersion and cut-off calculation are calculated by using CUDOS MOF Utilities [41] that simulates PCFs using the multipole method [42-43] along with MATLAB® for numerically calculating modes, the dispersion relation and cut-off parameter.

In this present study, we have shown that for the same values of normalized air-hole diameter, triangular-lattice PCFs show higher birefringence whereas the square-lattice one shows wider range of single-mode operation. We need smaller lengths of triangular PCF for dispersion compensation whereas PCFs with square-lattice can better compensate the broadband dispersion because of their near-matching dispersion slope with the existing inline fiber. The effective area of square-lattice is always greater than its triangular counterpart, making it better suitable for high power application.

## 2. Modal Properties
## 2.1 Birefringence properties:

The availability of large refractive index contrast between core and cladding, and flexibility to engineer the PCF geometrical parameters have made PCF a great candidate for achieving highly birefringent [4-15] fiber compared to conventional step-index fibers. Highly birefringent PCFs can be used as polarization maintaining fibers, which can stabilize the polarization states of the guided light.

Theoretically predicted birefringence for a series of fibers with different air-hole size is computed as a function of normalized frequency $\Lambda/\lambda$ and is shown in Fig. 2 (a) triangular-lattice air-holes (b) square-lattice air-holes. From Fig. 2, we can see that birefringence is strongly improving with the increase of air-hole size. From Fig. 2(a), we can see that the maximum birefringence of PCF with triangular-lattice air holes is $8.52\times10^{-3}$ at normalized frequency $\Lambda/\lambda$ =0.5 and $d/\Lambda$=0.7, which is almost 2 orders of magnitude higher that of the classical fiber. From Fig. 2(b), we can see that the maximum birefringence of PCF with square-lattice air holes is $6.03\times10^{-3}$ at normalized frequency $\Lambda/\lambda$ =0.486 and $d/\Lambda$=0.7. From the other values of $d/\Lambda$ that we have studied, it is observed that the maximum birefringence is obtained for PCF with triangular-lattice for the same $d/\Lambda$ value; whereas the maximum birefringence is available for a smaller value of normalized frequency $\Lambda/\lambda$ for PCF with square-lattice.

An interesting observation can be observed in the above figure for smaller values of air-filling fraction. For smaller values of $\Lambda/\lambda$, the birefringence reaches a minimum and then starts increasing again for further reduction of $\Lambda/\lambda$. As we go on decreasing the normalized frequency (higher wavelength), the refractive index difference between the two orthogonal modes decreases, subsequently reducing the birefringence. After a certain wavelength (equivalent $\Lambda/\lambda$) the effective indices of the two orthogonal modes crossover each other and after that they start to separate away from each other, thereby increasing the birefringence again. We should remember that we are actually calculating absolute difference of the effective indices (abs ($n_x$-$n_y$)) of the two orthogonal modes. So, the absolute birefringence reaches a minimum (at the cross over point) and then starts increasing again as can be observed with smaller $d/\Lambda$ values for both types of geometry.

Figure 3 shows the maximum birefringence variation with $d/\Lambda$ for PCFs with both triangular-lattice and square-lattice. We can clearly see that the birefringence is increasing with the increase of $d/\Lambda$. With the same value of $d/\Lambda$, PCFs with triangular-lattice are having higher values of birefringence than those of PCFs with square-lattice. The fact can be explained by the concept of air-filling fraction. For a triangular-lattice type of arrangement, the air-filling fraction is given by $f_\Delta = (\pi d^2)/(2\sqrt{3}\Lambda^2) = 0.9069(d^2/\Lambda^2)$, whereas for a square-lattice arrangement $f_\square = (\pi d^2)/(4\Lambda^2) = 0.7854(d^2/\Lambda^2)$.

From the two formulas, we can see that the air-filling fraction of PCFs with triangular-lattice has higher filling fraction than that of the PCFs with square-lattice. Therefore the higher value of air-filling fraction causes the PCFs with triangular-lattice to have higher value of birefringence.

Figure 4 shows the normalized frequency $\Lambda/\lambda$ corresponding to maximum birefringence as a function of $d/\Lambda$. We can see that the normalized frequency $\Lambda/\lambda$ corresponding to maximum birefringence decreases with the increase of $d/\Lambda$. At the same value of $d/\Lambda$, the normalized frequency $\Lambda/\lambda$ corresponding to maximum birefringence of PCFs with triangular-lattice is higher than that of PCFs with square-lattice.

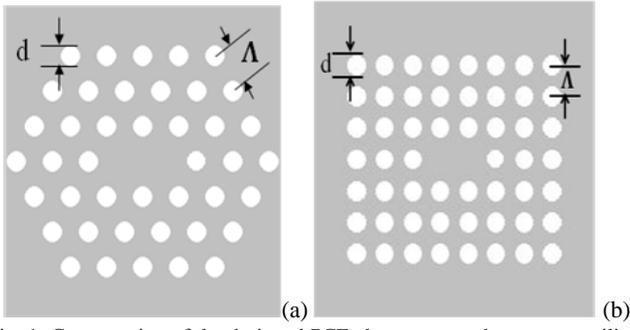

Fig. 1: Cross section of the designed PCF, the gray area denotes pure silica; the white area denotes air holes.

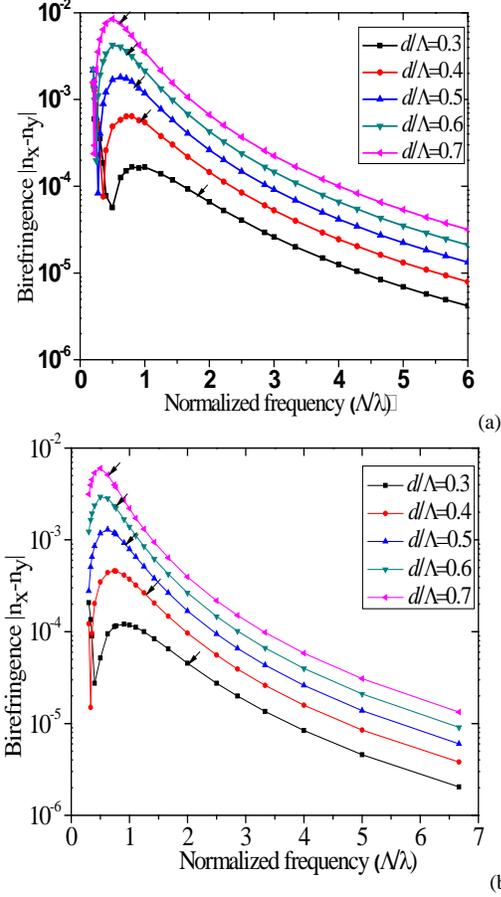

Fig. 2: Estimated birefringence as a function of $\Lambda/\lambda$ with $d/\Lambda$ as parameter. Arrow sign gives the transition between the first and second mode (a) triangular lattice PCF and (b) square lattice PCF.

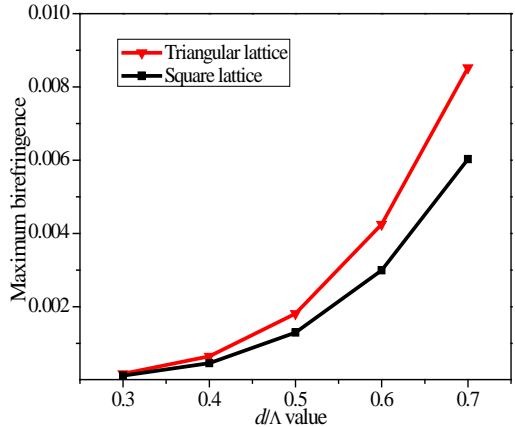

Fig. 3: Maximum birefringence as a function of $d/\Lambda$ for both types of PCFs.

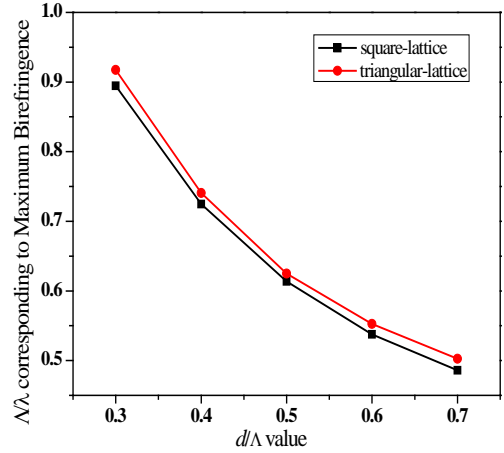

Fig. 4: Normalized frequency ($\Lambda/\lambda$) as a function of $d/\Lambda$ for both types of PCFs.

## 2.2: Cutoff properties and single mode region:

Triangular-lattice PCFs with a symmetric core are endlessly single mode for normalized hole sizes up to a value as large as $d/\Lambda=0.406$ [44-45], where as the symmetric core PCF with rectangular-lattice is endlessly single mode for normalized hole sizes up to a value as large as $d/\Lambda=0.442$[46]. This cannot be the case for the present fibers as the core region for the structure under concern is different from a normal one. We performed the cut-off analysis according to Kuhlmey *et al* [44] and Poli *et al* [46]. The multipole method [42-43] that has been used for the development of CUDOS MOF utilities has the unique ability to calculate both the modes and their losses accurately. In Fig. 5 we have shown the variation of the imaginary part of $n_{\text{eff}}$ (Im($neff$)) with $\lambda/\Lambda$. Im($neff$) is given by Eqn. (1)

$$\text{Im}(neff) = \text{Im}(\frac{\beta}{\kappa_0}) \qquad (1)$$

where $\beta$ is the complex propagation constant and $\kappa_0$ is the free-space wave number for the second mode. The figure represents Im($n_{\text{eff}}$) variation in a MOF with 3 rings (42 holes) for three different values of $d/\Lambda$, respectively 0.25, 0.30 and 0.40. Im($n_{\text{eff}}$) is linked to the geometrical losses $L$ (dB/km) through Eqn. (2) such that the loss and Im($n_{\text{eff}}$) are proportional at a fixed wavelength.

$$L = \frac{2\pi}{\lambda}\frac{20}{\ln(10)}10^9\,\text{Im}(n_{\text{eff}}) \qquad (2)$$

A clear shift from the regular variation can be observed as we increase the air-filling fraction ($d/\Lambda$). The figure shows a sharp transition in the ratio of Im($n$eff) (loss equivalent) versus $\lambda/\Lambda$ for $d/\Lambda>0.25$, whereas for $d/\Lambda\leq0.25$ the transition becomes increasingly gradual. With the increase of $d/\Lambda$ ($d/\Lambda=0.3$ and 0.4), the transition becomes more and more acute. With the increase of air-filling fraction, the curve deviated more and more from the original regular path signifying the PCF to be of multi-mode at higher $\lambda/\Lambda$ (or smaller $\Lambda/\lambda$).

The transition can be studied with some other parameters as well [44-46]. The transition can be better viewed if we plot the second derivative of the logarithm of the imaginary part of the effective index with respect to the wavelength, the Q parameter (specified by Eqn. (3)) as shown in Fig.6 [44].

$$Q = \frac{d^2 \log[\text{Im}(n_{eff})]}{d\lambda^2} \quad (3)$$

The approach has been a little different than which has been followed in their work. The wavelength has been changed in place of pitch for our study. The Q parameter shows a sharp negative minimum, giving an accurate value for the transition. Figure 6 clearly shows a distinct minimum for $d/\Lambda$ values of 0.30 and 0.40, whereas no minimum would be observed for $d/\Lambda$ value of 0.25. So whereas the regular PCF with triangular-lattice is single-mode up to a value of $d/\Lambda=0.4$, the present design is indeed found to support a second-order mode with a cut-off $\Lambda/\lambda=1.19$ for the same value of $d/\Lambda$. The fibers support second-order modes with a hole-size as small as $d/\Lambda=0.3$. We have studied the cut-off analysis of the PCF structure for triangular-lattice for $d/\Lambda$ values of 0.3, 0.4, 0.5, 0.6 and 0.7 and the second order cut-offs are marked with arrows in the Fig.2. Our study reveals that this fiber structure of PCF with triangular-lattice with hole-size $d/\Lambda=0.25$ or smaller may be classified as endlessly single-mode.

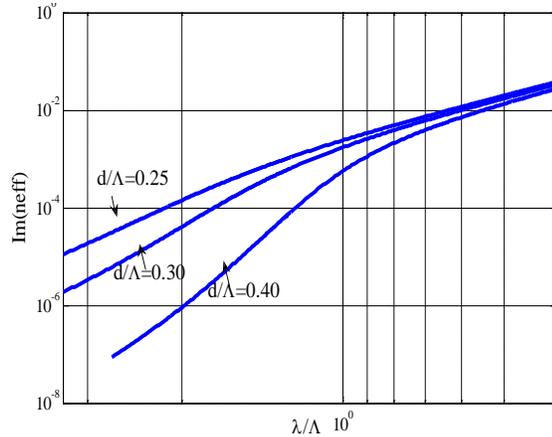

Fig. 5. Im($n_{eff}$) as a function of $\lambda/\Lambda$ for the PCF structure with triangular-lattice.

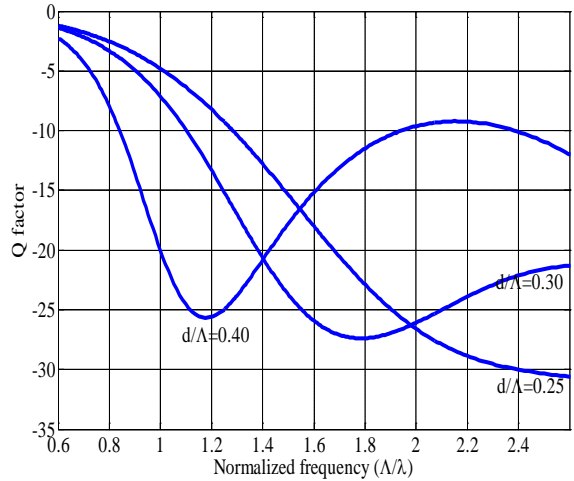

Fig. 6: Variation of the $Q$ factor (3) during the transition for the PCF with triangular-lattice.

With the same analogy as that of the PCF with triangular-lattice, we have studied the cut-off properties of the PCF with square-lattice as well. Figure 7 shows Im($n_{eff}$) of the second mode of the PCF with square-lattice. It is clearly visible that a transition is taking place for $d/\Lambda$ values greater than 0.28. Similarly to Fig. 5, with the increase of air-filling fraction, the loss curve deviated more and more from the original regular path signifying that the PCF, at higher $\lambda/\Lambda$ (or smaller $\Lambda/\lambda$), is becoming multi-mode. For an accurate determination of the transition point between the single-mode and multi-mode we have plotted the $Q$ parameter (Eqn. (3)) for the second mode of the PCF with square-lattice as shown in Fig (8). It is clearly shown that there is no minimum for the Q parameter for the $d/\Lambda$ value of 0.28.

Similar to the PCF with triangular-lattice, we have studied the cut-off analysis of the PCF with square-lattice for $d/\Lambda$ values of 0.3, 0.4, 0.5, 0.6 and 0.7 and the second mode cut-off regions are marked with arrows in the Fig.2 (b). Our study reveals that this fiber structure of PCF with square-lattice with hole-size $d/\Lambda=0.28$ or smaller may be classified as endlessly single-mode for PCFs with square-lattice.

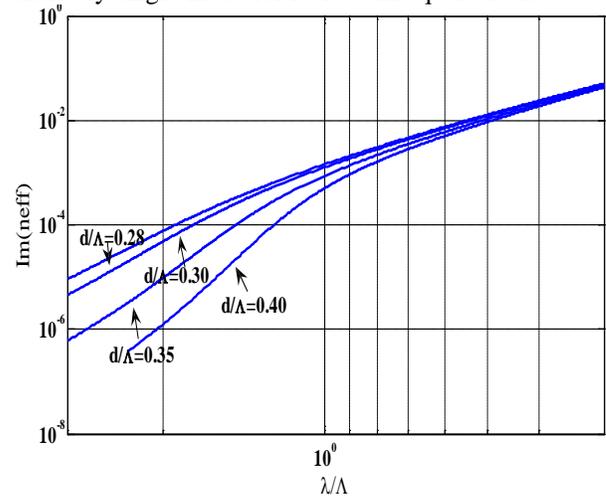

Fig. 7. Im($n_{eff}$) as a function of $\lambda/\Lambda$ for the PCF structure with square-lattice.

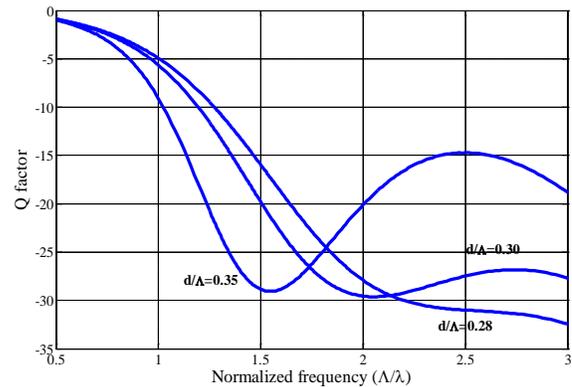

Fig. 8: Variation of the $Q$ factor (3) during the transition for the PCF with square-lattice.

Figure 9 shows the cut-off normalized frequency $\Lambda/\lambda$ as a function of $d/\Lambda$ for PCF with both triangular-lattice and square-lattice. We can see that the cut-off normalized frequency $\Lambda/\lambda$ for the PCFs decreases with the increase of $d/\Lambda$. At the same value of $d/\Lambda$, the normalized frequency $\Lambda/\lambda$ corresponding to maximum birefringence of PCFs with triangular-lattice is lower than that of PCFs with square-lattice. So PCF with square-lattice is single mode for higher values of $d/\Lambda$.

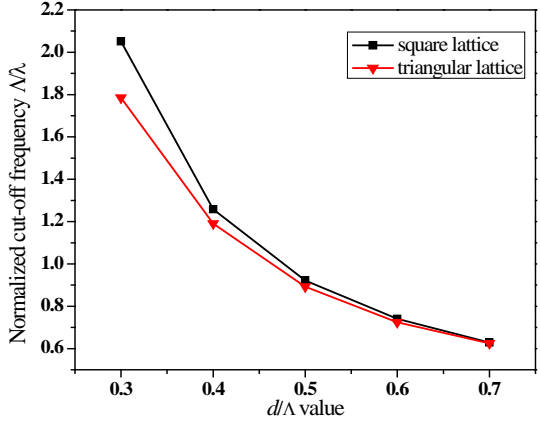

Fig. 9: Normalized cut-off frequency $\Lambda/\lambda$ as a function of $d/\Lambda$ for both types of PCFs.

### 2.3: Dispersion properties:

PCFs possess the attractive property of great controllability in waveguide dispersion [18]. The dispersion property for both the PCFs with triangular-lattice and square-lattice can be controlled by varying the air-hole diameter $d$ and pitch $\Lambda$. Controllability of waveguide dispersion in PCFs is an important issue for applications to optical communications, dispersion compensation, non-linear applications, etc. We have calculated the dispersion of the structures through Eqn. (4).

$$D = -\frac{\lambda}{c}\frac{d^2 n_{eff}}{d\lambda^2} \quad (4)$$

Figure 10 shows the dispersion properties of the PCF with triangular-lattice and square-lattice for x-polarized component for different values of air-filling fraction ($d/\Lambda$) with $\Lambda=1\mu m$. It can be observed that for both types of fibers dispersion slope is always positive for smaller values of air-filling fraction (0.3 and 0.4). With the increase of $d/\Lambda$ value to 0.5 the dispersion values change drastically and the slope is always negative throughout the wavelength range we considered. As we go still further for higher values of $d/\Lambda$, though the slope is negative the effective dispersion values become more and more positive. So both types of fiber with moderate value of air-filling fraction (0.5 here) can better compensate the dispersion. When $\Lambda$ becomes larger the effect of waveguide dispersion decreases and material dispersion dominates the dispersion for both regular triangular-lattice PCFs [18, 47] and for square-lattice PCF [38-39]. The same is confirmed for both types of structures as shown in Fig. 11 where we have considered higher values of $\Lambda$ for both elliptical core triangular-lattice and square-lattice PCFs in Fig. 11(a) and 11(b), respectively, as we increase the $\Lambda$ to $2\mu m$. Figure 11 shows the dispersion properties of the structures for x polarized component for higher values of $\Lambda$ ($\Lambda=2\mu m$). When the $d/\Lambda$ is very small and $\Lambda$ is large, the dispersion curve is close to that of the material dispersion of pure silica. As the air-hole diameter is increased, the influence of waveguide dispersion becomes stronger. We can see that it is possible to shift the zero dispersion wavelength from visible to near-infrared (IR) by appropriately changing the geometrical parameters such as $\Lambda$ and $d$ for both type of PCFs with triangular-lattice and square-lattice. For even higher values of $\Lambda$ (=3$\mu$m, not shown here) the dispersion values are all positive with a higher positive slope than with $\Lambda$ =2$\mu$m. A comparison of the two types of PCFs with higher $\Lambda$ (=2.5$\mu$m) and $d/\Lambda$ =0.4 has been considered for a wideband wavelength range as shown in Fig. 12. The two dotted lines show the ZDW corresponding to the two different PCF structures. The graph establishes some significant outcomes as explained below. First of all, with the similar geometrical parameters, the ZDW of the square-lattice PCF is red-shifted which is preferable for mid-IR SCG, especially those using non-silica high nonlinear chalcogenide PCFs. Secondly, the dispersion profile of the square-lattice PCF shows a higher slope (*i.e.* the curve is less flat) than the dispersion characteristics revealed by similar type regular triangular-lattice structures. The high slope in the dispersion curve leads to compression of the generated broadband accumulating more power in the supercontinuum pulse. On the other hand, with flatter slope with triangular PCF, we could have a wider range of SCG. Thirdly, square-lattice PCF has low GVD at the anomalous dispersion region which corresponds to higher Dispersion Length ($L_D$) and higher degree of solitonic interaction. The effective area corresponding to the above structure has been presented in Fig. 13 which clearly shows that square-lattice PCF has a higher amount of effective area which is preferable for high power applications as the larger core can accumulate a higher amount of power. Also higher effective area means a higher threshold power limit before material damage occurs.

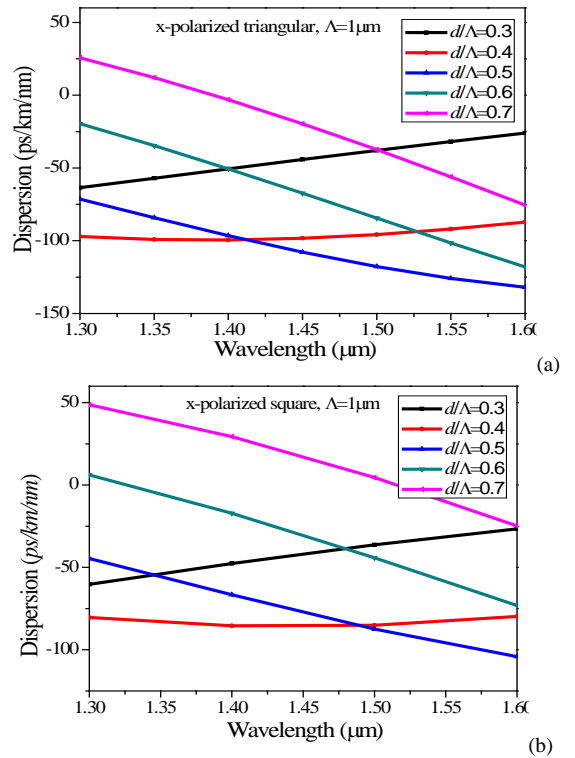

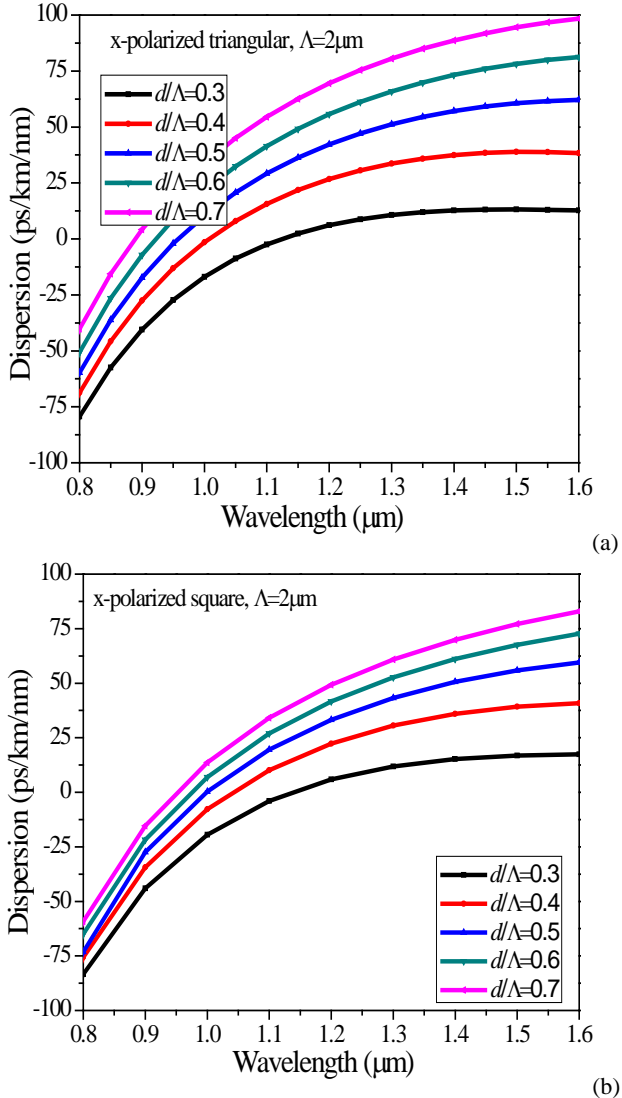

Fig. 10: Comparison of the *D* parameter for both types of PCFs with different value of *d*/Λ with Λ=1μm for (a) triangular-lattice PCF and (b) square-lattice PCF.

Fig. 11: Comparison of the dispersion parameter values *D* for both the polarization for both types of PCFs with Λ=2μm for (a) triangular-lattice PCF and (b) square-lattice PCF.

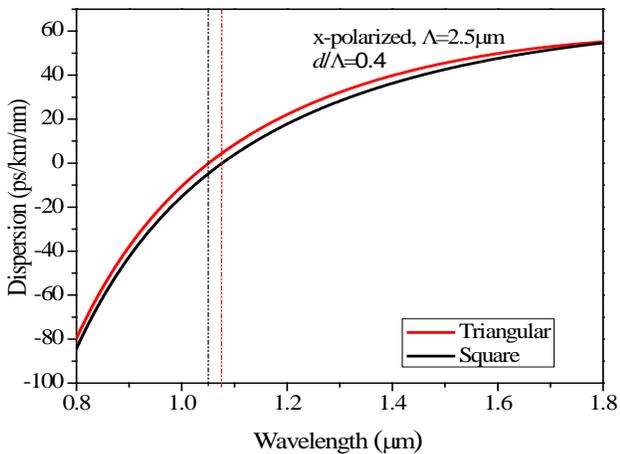

Fig. 12: Comparison of the dispersion properties (*x*-polarization) of triangular-lattice and square-lattice PCFs with *d*/Λ =0.4, for Λ=2.5μm. ZDWs for both the structures are shown by the dotted lines.

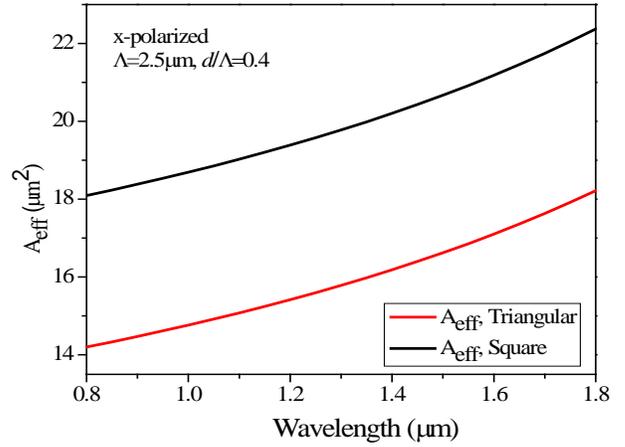

Fig. 13: Comparison of the effective area variation of both types of PCFs keeping Λ=2.5μm with *d*/Λ=0.4.

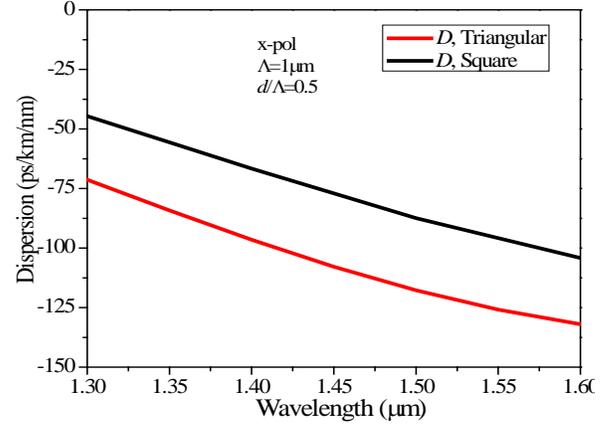

Fig. 14: Comparison of the dispersion variation for both types of PCFs keeping Λ=1μm with *d*/Λ=0.5.

A comparison for dispersion compensation has been carried out in Fig. 14, where we have taken Λ=1μm and *d*/Λ=0.5. The graph clearly indicates that both types of fibers can be useful for dispersion compensation while higher values of negative dispersion can be achieved with the triangular one. However, the relative dispersion slope (RDS) value, which is very important for broadband dispersion compensation [48], is better for the square-lattice with RDS of 0.00175nm$^{-1}$, whereas for the triangular one the value is 0.00113 nm$^{-1}$ at 1550nm with the RDS value of the standard SMF-28 at this wavelength is 0.0036 nm$^{-1}$.

### 2.4: Single mode properties comparison:

A final analysis on the properties of both types of elliptical core PCFs is reported in Fig. 15 for different values of Λ values namely 1μm and 2μm. We have taken *d*/Λ values to be 0.3 for our study such that both the elliptical core PCFs remains single-mode in the whole wavelength considered. As we have already observed that elliptical core square-lattice is endlessly single mode up-to a *d*/Λ value of 0.28, where as the elliptical core triangular PCFs are endlessly single-mode for *d*/Λ value up-to 0.25. We have observed that cut-off normalized frequency (Λ/λ) for elliptical core triangular-lattice PCF and square–lattice PCF are 1.785 and 2.041, respectively for *d*/Λ=0.3. Taking into consideration of the above values we have restricted our study for single mode operation. Square-lattice PCF has higher *D* values than triangular PCF for larger values of Λ(*i.e.* 2μm) in the wavelength range considered. An interesting fact can be observed from the figure that square-lattice PCF has higher *D* values than triangular PCF for smaller values of Λ(*i.e.* 1μm) but the valu**es** cross-over and reverse phenomenon took place after cross-over. The dispersion slope has been affected a little with the geometrical characteristics of the lattice. With the increase of the Λ values the slope almost vanishes as we move towards the higher wavelength ranges. PCFs with square-lattice always have higher effective area than triangular one for both the pitches as shown in Fig. 15(b).

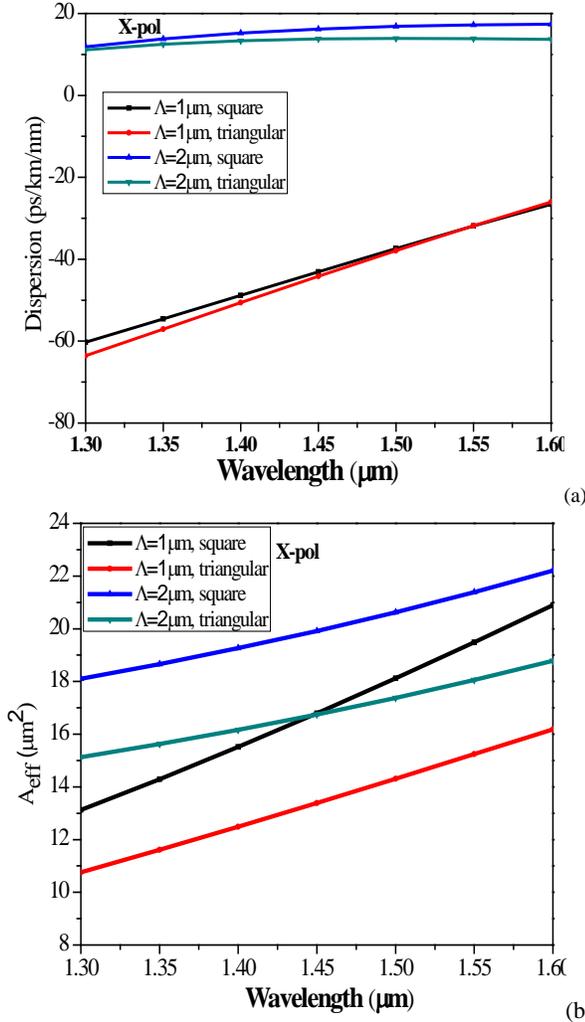

Fig. 15: Comparison of (a) dispersion variation (b) effective area of the two types of PCFs when they are under single mode operation.

## 2.5: Comparison with symmetrical-core PCF

We have performed a comparative analysis of dispersions properties of the triangular lattice PCF between the symmetric-core and the asymmetric-core geometry. As we have already observed that the best performance in terms of broadband dispersion compensation with elliptical core PCF can be achieved with Λ=1μm and *d*/Λ=0.5 (Fig. 10), we have considered the above parameters (Λ= 1μm with *d*/Λ=0.5) for both the cores to have a comparison with the regular triangular-lattice PCF. The findings are demonstrated in Fig. 16. It can be observed from the figure that regular symmetric-core PCF has higher negative value of dispersion at the communication wavelength of 1550nm, which clearly indicates that we need smaller length of the fiber for total dispersion compensation. However, the slope of the dispersion curve for symmetric-core PCF around 1550nm of wavelength is positive. It is well known that dispersion accumulates due to propagation of light in the existing inline fiber (SMF-28 is) is positive with a positive slope. Therefore, to compensate the broadband dispersion, our dispersion compensating fiber should consist of negative dispersion with a negative slope. So, broadband dispersion compensation can't be possible with the above geometrical parameter with symmetric-core PCF. Consequently, for broadband dispersion compensation, elliptical-core PCF with negative dispersion and negative slope with Λ=1μm and *d*/Λ=0.5 will be a better choice.

In another comparison, we have considered higher Λ (=2.5μm) and smaller *d*/Λ(=0.4) for both types of symmetric-core and asymmetric-core PCFs. The comparative study has been demonstrated in Fig. 17 with the value of the ZDW presented with the vertical dotted line. The figure clearly presents that the ZDW of the asymmetric-core PCF is red-shifted compared to that of symmetric-core PCF, which in turn establishes that mid-IR SCG can be better performed with elliptical core PCF with the same geometrical parameters.

It is also interesting to note that the dispersion profile of the asymmetric-core PCF shows a higher slope (*i.e.* the curve is less flat) than the dispersion characteristics revealed by similar type regular symmetric-core PCF structures. The higher slope in the dispersion curve leads to compression of the generated broadband accumulating more power in the supercontinuum pulse. On the other hand, with flatter slope with symmetric-core PCF, we could have a wider range of SCG.

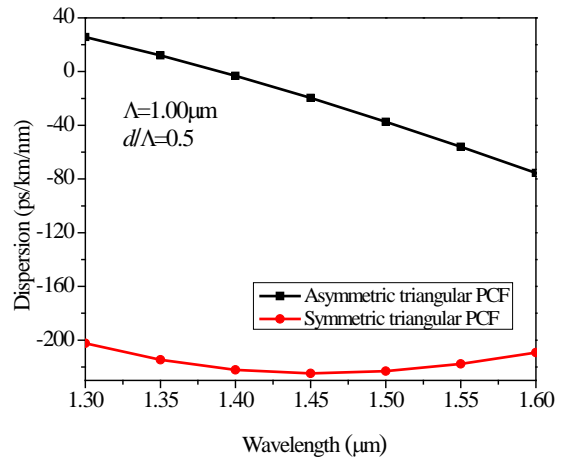

Fig. 16: Comparison of the dispersion properties for symmetric-core and asymmetric-core PCFs with Λ=1μm and *d*/Λ=0.5.

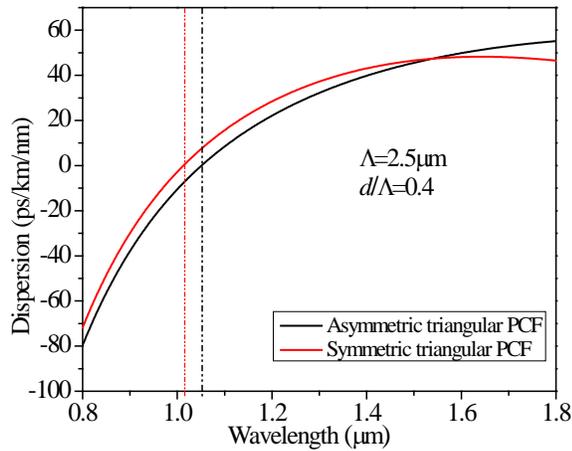

Fig. 17: Comparison of the dispersion properties for symmetric-core and asymmetric-core PCFs with Λ=2.5μm and $d/\Lambda$=0.4. ZDWs for both the structures are shown by the dotted lines.

3. Conclusion:

We have shown that a high birefringence of the order of $10^{-2}$ can be achieved in both the PCFs with triangular-lattice and square-lattice air-hole arrangement in the cladding and a silica core of the PCFs that is formed by removing two adjacent holes in the center of the fiber. Our numerical calculations establish that maximum birefringence can be achieved with PCFs with triangular-lattice, where as wider range of single mode operation is possible for PCFs with square-lattice. Triangular-lattice PCF was found to be endlessly single-mode for an air-filling fraction ($d/\Lambda$) of 0.25, whereas the square-lattice PCF was found to be endlessly single-mode for higher $d/\Lambda$ of 0.28. A comparison regarding the dispersion characteristics of the two structures reveal that, we need smaller length of triangular-lattice PCF for dispersion compensation, whereas PCFs with square-lattice can better compensate broadband dispersion compensation due to the better matching of the *RDS* with the existing SMF-28. In addition to the above, square-lattice PCFs show ZDW red-shifted making it preferable for mid-IR SCG with chalcogenide material. Also its dispersion profile with higher dispersion slope leads to compression of the broadband accumulating more power in the pulse. On the other hand, triangular-lattice PCF with flat dispersion profile can generate broader SCG. Square-lattice PCF with low GVD at the anomalous dispersion corresponds to higher dispersion length ($L_D$) and higher degree of solitonic interaction. The effective area of square-lattice PCF is always greater than its triangular-lattice counterpart making it better suitable for high power accumulations and enhanced threshold power limit for material damage. We have also performed a comparison of the dispersion properties of between the symmetric-core and asymmetric-core triangular lattice PCF. While we need smaller length of symmetric-core PCF for dispersion compensation, broadband dispersion compensation can be performed with asymmetric-core PCF with Λ=1μm and $d/\Lambda$=0.5. Mid-IR SCG can be better performed with asymmetric core PCF with compressed broadband that can accumulate high power in the pulse because of higher slope, while on the other hand symmetric core PCF can generate a wider range of SCG because of its flatter slope. Thus, these analyses will be extremely useful for realizing fiber aiming towards a custom application around these characteristics.